\documentclass{article}

\PassOptionsToPackage{numbers, compress}{natbib}

\usepackage[preprint]{neurips_2024}

\usepackage[utf8]{inputenc} 
\usepackage[T1]{fontenc}    
\usepackage{hyperref}       
\usepackage{url}            
\usepackage{booktabs}       
\usepackage{amsfonts}       
\usepackage{nicefrac}       
\usepackage{microtype}      
\usepackage{xcolor}         
\usepackage{multirow}
\usepackage{multicol}
\usepackage{algorithm}
\usepackage{algpseudocode}
\usepackage{graphicx}

\newcommand{\LEARNANYWAY}{\textsc{LearnAnyWay}}

\title{LLMs as Probabilistic Minimally Adequate Teachers for DFA Learning}

\author{%
  Lekai Chen\thanks{Use footnote for providing further information
    about author (webpage, alternative address)---\emph{not} for acknowledging
    funding agencies.} \\
  Department of Computer Science\\
  University of Colorado, Boulder\\
  Boulder, CO 80309 \\
  \texttt{Lekai.Chen@colorado.edu} \\
  \And
  Ashutosh Trivedi \\
  Department of Computer Science\\
  University of Colorado, Boulder\\
  Boulder, CO 80309 \\
  \texttt{Ashutosh.Trivedi@colorado.edu} \\
  \AND
  Alvaro Velasquez \\
  Department of Computer Science\\
  University of Colorado, Boulder\\
  Boulder, CO 80309 \\
  \texttt{Alvaro.Velasquez@colorado.edu} \\
}

\begin{document}

\maketitle

\begin{abstract}
The emergence of intelligence in large language models (LLMs) has inspired investigations into their integration into automata learning. 
This paper introduces the probabilistic Minimally Adequate Teacher (pMAT) formulation, which leverages a probabilistic oracle that could give persistent errors randomly during answering the membership queries for deterministic finite automata (DFA) learning. 
Given the tendency of LLMs to produce hallucinatory content, we have developed techniques to improve answer accuracy and ensure the correctness of the learned automata. 
We propose the $\mathtt{Discrimination}$ prompt as well as the $\mathtt{Verification}$ prompt and explore their advantages over common prompts. 
Additionally, we compare DFA learning performance between the TTT algorithm and common active learning algorithms. 
To address the exponential number of persistent errors, we implement a dynamic query cache refinement algorithm that identifies and corrects conflicting queries by combining the active and passive learning algorithms. 
The empirical results demonstrate the robustness and efficiency of our approach, providing a theoretical foundation for automata learning with LLMs in the loop.    
\end{abstract}

\section{Introduction}

Automata serve as fundamental constructs of encoding knowledge and rules in deep learning area, such as reward machine in reinforcement learning\cite{NEURIPS2019_532435c4, Toro_Icarte_2022}, sequence modeling \cite{bejerano2003automata}, and interpretable neural networks\cite{liu2023transformers, merrill-2019-sequential}. With the emergence of LLMs, they are exploited as rich sources of human knowledge. LLMs are also valued for their reasoning and planning abilities\cite{xi2023rise}. However, directly using the text-based responses for logic inference and decision making are often challenging. 
Therefore, LLMs are trained to answer in more regular and structured ways, such as mathematical expressions\cite{imani2023mathprompter, meadows2023generating, Romera-Paredes2024}, codes\cite{chen2021evaluating, nijkamp2023codegen}, and structured data (JSON)\cite{Dagdelen2024}. 

\begin{figure}[t]
  \centering
  \includegraphics[width=0.8\linewidth]{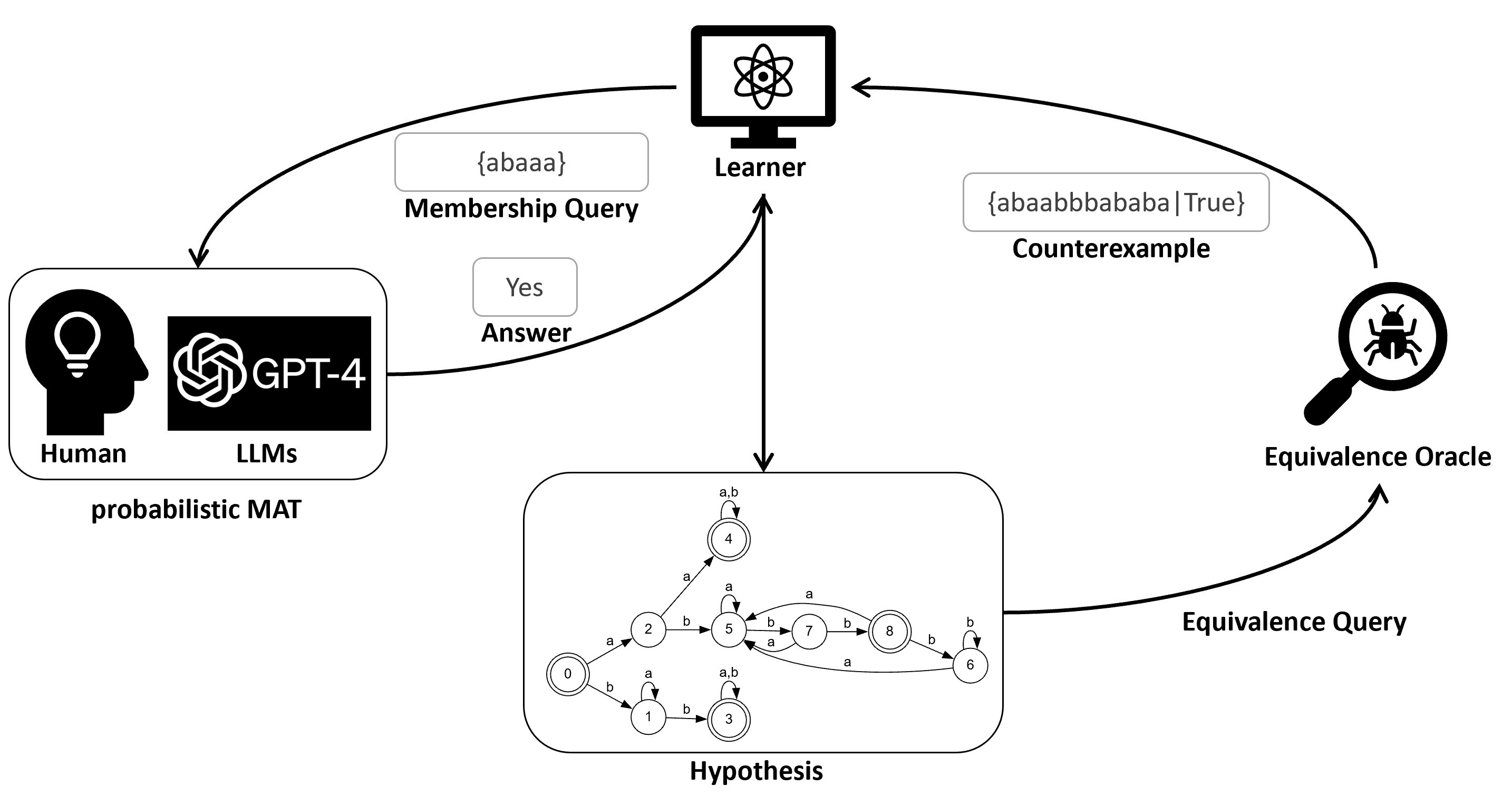}
  \caption{Probabilistic minimally adequate teacher formulation with LLMs or human in the loop.}
  \label{fig:pMAT}
\end{figure}

While useful in specific contexts like mathematics and programming, these approaches often fall short in capturing formal logical structures. This limitation has led to explorations into using LLMs for automata learning, where LLMs respond to answer membership queries\cite{vazquezchanlatte2024llm} or directly output transitions\cite{alsadat2024using}. However, the probabilistic nature of LLMs makes them unreliable oracles, incapable of consistently producing correct responses, thus complicating the accurate construction of DFAs. None of the methods above can promise to construct a DFA equivalent to the ground-truth model, because they fail at detecting and correcting the wrong response from LLMs. To address this, we introduce the probabilistic Minimally Adequate Teacher (pMAT), shown in figure \ref{fig:pMAT}, where the oracle may inaccurately answer membership queries but will always provide valid counterexamples to aid in refining hypotheses. We use this formulation for 3 reasons: 
1) LLMs are adept at handling membership queries but struggle with equivalence queries\cite{vazquezchanlatte2024llm}; 
2) LLMs cannot be invariably trusted. The errors produced by LLMs are persistent and cannot be corrected by repetitive sampling; and 
3) Practical implementations often validate hypotheses directly against the target model, such as in software or environmental simulations, making the valid counterexamples available.

To increase the accuracy of membership queries answering, we leverage prompt engineering which is a common solution to enhance the LLM's reasoning ability. One of the most effective and reliable prompt method is Chain of Thought\cite{NEURIPS2022_9d560961}, which uses a series of intermediate reasoning steps to teach LLMs how to solve the problem step by step. Inspired by CoT and specificity of our task, we develop two novel prompting strategies, $\mathtt{Discrimination}$ prompt and $\mathtt{Verification}$ prompt. The prefix and suffix prompt can divide a long word in to a prefix and a suffix such that the length of the membership query can be much lower. It also allows LLMs to utilize the results of previous queries. Concurrently, the $\mathtt{Verification}$ prompt, deriving from CoT principles, compels LLMs to substantiate their analyses, fostering more reliable and correct reasoning.

Another strategy to construct a correct automaton involves minimizing the number of membership queries by advanced automata learning algorithms. The stochastic nature of LLM outputs implies that increased querying correlates with more errors. Unfortunately, if a learner is given any wrong membership query, it cannot build a correct automaton, because the counterexample will conflict with erroneous queries. They cannot be solved by simply sampling more LLM responses, because LLMs tend to produce the same results for one single query. Therefore, we minimize query reliance by implementing the TTT algorithm \cite{10.1007/978-3-319-11164-3_26}, which is renowned for its efficiency in analyzing counterexamples and requires fewer membership queries compared to methods like $L^*$~\cite{ANGLUIN198787, KV}.

Despite these enhancements, the inherent inaccuracy of LLM responses means that a perfect learning outcome cannot be guaranteed solely through reduced querying and improved prompts. Angluin and Kriķis proved that DFAs can be learned with finite exceptions given a reliable equivalence oracle, and they proposed the algorithm \LEARNANYWAY{} to handle erroneous membership queries by simply replacing them with labeled counterexamples \cite{Angluin1997}. However, in a runtime verification environment, the shortest optimal counterexamples are often not available \cite{10.1007/978-3-319-11164-3_26}. As the average length of counterexamples increases, the number of errors introduced by the active learner during hypothesis refinement grows exponentially, rendering the \LEARNANYWAY{}  algorithm infeasible due to the massive number of LLM and equivalence queries required, even though the exceptions remain finite.

To address this issue, we introduce \LEARNANYWAY{}  with passive refinement (LAPR), which combines active and passive learning algorithms to efficiently eliminate persistent errors. This algorithm creates a cache to maintain the membership queries corresponding to each counterexample. Upon detecting conflicts between a counterexample and cached queries, these queries are reassessed and their labels adjusted based on a passive learner's hypothesis trained on MQs and EQs, enabling continuous hypothesis refinement. This approach ensures the progressive learning of an automaton that is equivalent to the target model, as demonstrated by our empirical validations.

\section{Related Works}
\paragraph{Active Learning Algorithms and the MAT Framework.}
Active learning algorithms \cite{ANGLUIN198787, KV, rivest1989inference, Parekh1997APT, denis2004learning, bongard2005active, volpato2015approximate, 10.1007/978-3-319-11164-3_26}, such as the LStar, are pivotal in the domain of DFA learning. These algorithms keep asking an oracle, which provides answers to MQs and EQs, to iteratively refine a hypothesis automaton until it accurately constructs the target automaton. The concept of a Minimally Adequate Teacher (MAT) introduced by Angluin is central to this process \cite{ANGLUIN198787}. A MAT is an oracle that can answer both MQs—determining whether a given string is part of the target language—and EQs—providing counterexamples when the current hypothesis does not match the target automaton.

In practice, oracles may not always provide correct answers to MQs and EQs. \citet{moeller2023automata} propose a new formulation called the incomplete Minimally Adequate Teacher (iMAT), which addresses scenarios where the teacher has access to only a finite number of tests or has gaps in its knowledge. The iMAT framework can still provide answers that are correct within the limited scope of its knowledge, a guarantee that LLMs cannot offer.
\citet{Angluin1997} delve into the issues of errors or omissions in membership query responses and the learning of finite variants of concepts in polynomial-time exact learning using membership and equivalence queries. They demonstrate that the class of regular languages, such as DFAs, is learnable in polynomial time with equivalence and malicious membership queries. However, their approach becomes impractical when dealing with the exponential increase in errors in MQs as the average length of the counterexamples grows.

\paragraph{Passive Learning Algorithms.}
In contrast, passive learning algorithms like RPNI (Regular Positive and Negative Inference) and its variants (e.g., RPNI-EDSM) infer a DFA from a fixed set of positive and negative examples without interactively querying an oracle. Instead, they rely on state merging \cite{doi:10.1142/9789812797919_0007, lang1998results}. These algorithms construct an initial hypothesis from both the positive and negative examples and refine it to accept all given examples, often achieving a model that generalizes well to unseen data. However, the learning results are often incomplete due to the limited scope of the provided samples and the potential for overfitting \cite{bugalho2005inference}. \citet{yang2019improving} propose a hybrid approach that combines the strengths of active and passive learning algorithms to improve automata learning. They improve L-star algorithm by integrating execution logs and results of passive learning. However, this approach does not address the issue of incorrect membership queries. When an oracle provides erroneous responses, the hybrid approach lacks mechanisms to detect and correct these errors.

\paragraph{Using LLMs as Oracles.}
Recent advancements have explored the use of Large Language Models (LLMs) as oracles in active learning for automata. \citet{vazquezchanlatte2024llm} employ the LStar algorithm, leveraging LLMs to answer membership queries. They enhance the query answer ability of LLMs by allowing the models to say unsure and subsequently use the DISS search algorithm to find answers. However, due to the probabilistic nature of LLMs, their responses may contain persistent errors, undermining the guarantee of learning an automaton equivalent to the target model. \citet{alsadat2024using} utilize LLMs to translate natural language knowledge into propositions and transitions of automata. This method refines the learned DFA by utilizing counterexamples collected from the environment. However, this approach also lacks a guarantee of correctness since it does not employ existing DFA learning algorithms, and the learned DFA is merely used as a reward machine to expedite reinforcement learning optimization.

\section{Preliminaries}

\paragraph{Deterministic Finite Automata (DFA).}
The DFA is a common variant of regular languages. Here, we will introduce the basic concepts and notations of a DFA. Let $\Sigma$ be a finite alphabet. Consider a finite set of symbols, $\Sigma$, known as the alphabet. A DFA, denoted as $\mathcal{A}$, over $\Sigma$ is defined as a quintuple $\mathcal{A} = \langle Q^{\mathcal{A}}, \Sigma, q_0^{\mathcal{A}}, \delta^{\mathcal{A}}, F^{\mathcal{A}} \rangle$, where:
$Q^{\mathcal{A}}$ is a finite set of states; 
$\Sigma$ is a pre-defined finite alphabet; 
$q_0^{\mathcal{A}} \in Q^{\mathcal{A}}$ is the initial state; 
$\delta^{\mathcal{A}}: Q^{\mathcal{A}} \times \Sigma \to Q^{\mathcal{A}}$ is the transition function; and
$F^{\mathcal{A}} \subseteq Q^{\mathcal{A}}$ is the set of final (or accepting) states.

For a symbol $a \in \Sigma$ and a state $q \in Q^{\mathcal{A}}$, the state $q' = \delta^{\mathcal{A}}(q, a)$ is referred to as the $a$-successor of $q$. The transition function extends to words where $\delta^{\mathcal{A}}(q, \varepsilon) = q$ and $\delta^{\mathcal{A}}(q, wa) = \delta^{\mathcal{A}}(\delta^{\mathcal{A}}(q, w), a)$ for any $q \in Q$, $a \in \Sigma$, and $w \in \Sigma^*$. To evaluate a word $v \in \Sigma^*$ under a DFA, we define the output function as $\mathcal{A}(q): \Sigma^* \to {true, false}$, such that $\mathcal{A}(q, v) = true$ iff $\delta^{\mathcal{A}}(q, v) \in F^{\mathcal{A}}$. The function $\mathcal{A}[u] = \delta^{\mathcal{A}}(q_0^{\mathcal{A}}, u)$ maps a word $u \in \Sigma^*$ to the state reached by $u$. A query is a tuple $(w, [yes,no,unknown])$ where $w\in \Sigma^*$ is the word to be queried. To represent the cache that stores membership queries and equivalence queries, we use $C_{MQ}$ and $C_{EQ}$. The access to result of a query $x$ in the cache is denoted by $C[x]$.

\paragraph{Minimally Adequate Teacher (MAT).}
The MAT framework involves a Learner ($L$) attempting to deduce an unknown language by interacting with a Teacher, who responds to two types of queries: Membership Query (MQ) and Equivalence Query (EQ). In an MQ, the Learner proposes a word $u$, and the Teacher confirms if $u$ is part of the language. In an EQ, the Learner offers a hypothetical automaton $\mathcal{H}$ to the Teacher, who either verifies $\mathcal{H}$ or provides a counterexample if incorrect.

\paragraph{The pMAT Formulation.}
We introduce a new formulation for DFA learning called the probabilistic Minimally Adequate Teacher (pMAT), which can err in responding to membership queries with a probability $\epsilon$, though counterexamples remain accurate. Similarly, the automata learner in pMAT aims to learn a target automaton based on partially correct MQ responses and correct EQ responses. The target automaton can be learned solely depending on the counterexamples because they are the only verified answers.

The pMAT formulation, See Figure \ref{fig:pMAT}, consists of three components: the automata Learner $L$, the probabilistic membership oracle $O_{MQ}$, and the equivalence oracle $O_{EQ}$. The Learner refines its hypothesis based on membership queries and requests for counterexamples. Membership queries involve checking the presence of a word $w$ in the target language, whereas equivalence queries assess if the hypothesis $\mathcal{H}$ matches the target model, with a counterexample provided if it does not.

The probabilistic nature of the Teacher in pMAT implies that errors may occur independently and at a constant probability $\epsilon$ for new membership queries, simulated by a binary distribution $P$. Once an error occurs, it will persist and cannot be corrected by multiple samplings, as there is no guarantee that the more frequent response is more reliable. For example, consider taking $N$ samples from $P$, where the probability of an erroneous response is $\epsilon$, but the specifics of which samples are erroneous remain unknown. In contexts like Large Language Models (LLMs), which operate on vast datasets and generally provide accurate responses, errors are rare but can occur, making every query's outcome potentially significant. This necessitates caching query results to avoid losing data from repeated queries, thereby making any errors persistent.

\section{Methods}

\paragraph{Verification Prompt.}
The $\mathtt{Verification}$ prompt extends the capabilities of the Chain of Thought (CoT) prompts. Initially, Large Language Models (LLMs) respond to membership queries by performing analyses similar to those demonstrated in example queries as we see in figure \ref{fig:veri_prompt}. However, due to the tendency of LLMs to generate hallucinated content, these analyses may not always be relevant to the actual query input or the language definitions, leading to incorrect membership decisions. To address this issue, we introduce a $\mathtt{Verification}$ step.

Upon receiving the initial response from the LLM to a CoT query, a secondary process begins where a "teacher" LLM checks whether the response aligns with the expected definitions and the context of the query. This involves a comparison of the initial analysis against a set of language definitions that constitute a correct response, as well as the original query input. The LLMs are required to output a JSON object containing three components:
\begin{itemize}
    \item Whether the analysis matches the language definition,
    \item Whether the analysis is consistent with the query input,
    \item The reasons why the analysis matches the definitions, as well as why the analysis is based on the query input.
\end{itemize}

If the teacher finds discrepancies between the initial answer and the expected standards, the following actions are taken: a) If the input sequence and definition match but the initial response was incorrect, the answer is inverted (changed from true to false, or vice versa). b) If there is a mismatch in the input sequence or definition, the system revises its additional prompt based on the teacher's analysis and re-evaluates the query. This verification step significantly enhances the reliability of the system's outputs.

By ensuring that each membership query's response is not only generated but also rigorously checked, the LLMs minimize errors and align their outputs closely with accurate interpretations of the language rules.

\begin{figure}[t]
  \centering
  \includegraphics[width=\linewidth]{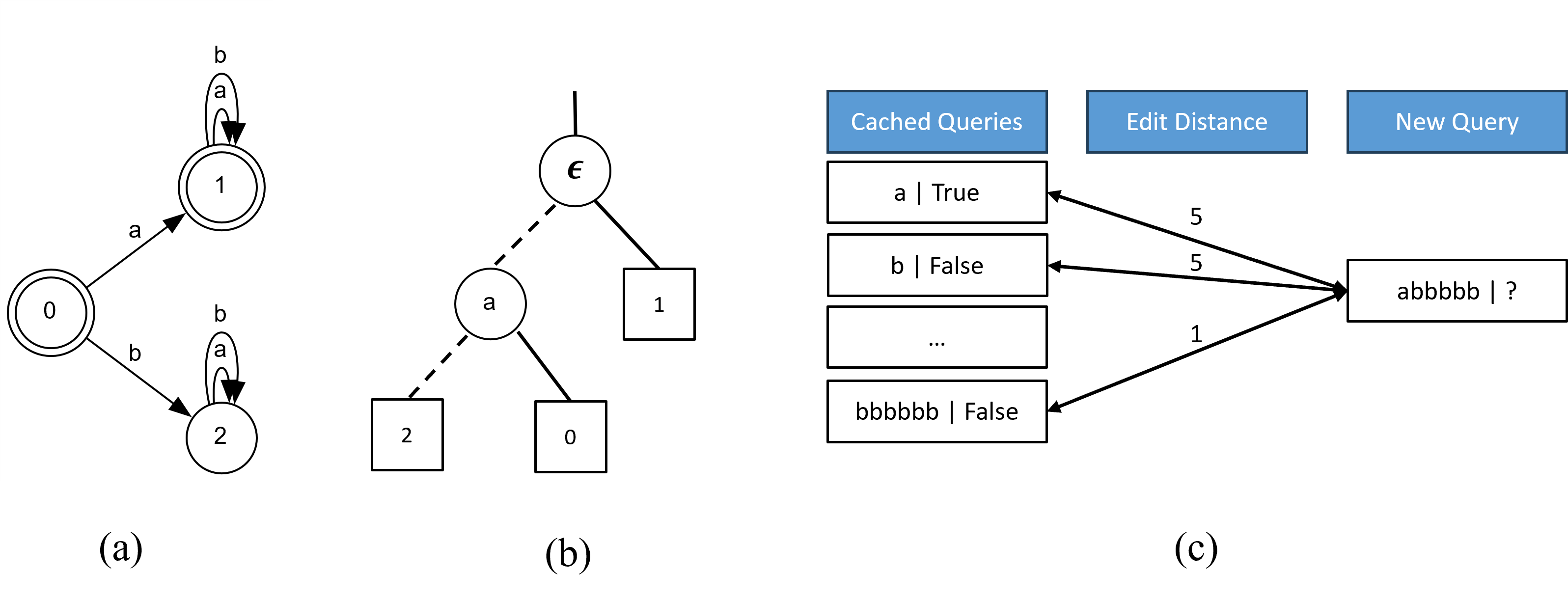}
  \caption{Discrimination prompt running example: (a) target automaton that only accepts the string starting with $a$. (b) The corresponding discrimination tree to the target DFA. The leaves are states in the automaton. The inner nodes represent the discriminator that makes the states in left and right side different. (c) The edit distance between the new query and the cached queries.}
  \label{fig:disc_prompt}
\end{figure}

\paragraph{Discrimination Prompt.}

The idea behind the $\mathtt{Discrimination}$ Prompt is that LLMs perform better with examples compared to direct inference. However, handling the large volume of membership queries necessary for learning an automaton is challenging, particularly when it's impractical to present all cached queries to LLMs. That is, we have to find the most similar ones to prompt LLMs. Intuitively, words that appear similar should share properties within the same language, but this isn't universally applicable. See the example in figure \ref{fig:disc_prompt}. Consider an automaton in (a), for a membership query on the word $\langle abbbbb \rangle$, among the cached queries in (c). The $\langle bbbbbb \rangle$ appears more similar due to a lower edit distance. Yet, prompting LLMs with $\langle bbbbbb \rangle$ might lead them to incorrectly infer non-membership, mirroring its negative example. Conversely, using $\langle a \rangle$ as a prompt highlights that valid strings should start with 'a', though it might not instill confidence due to significant differences from the query word.

To address these challenges, we implement the $\mathtt{Discrimination}$ prompt, which maintains a discrimination tree in (b) to remember the equivalence queries and their relationships. Upon receiving a new membership query, the algorithm identifies the query's position within the tree—specifically, which leaf (state) the new word belongs to. It then proceeds to the lowest common ancestor of that leaf and selects the most similar word by edit distance from each child of that ancestor. Words on the same leaf share the same state in the hypothesized automaton, while words on adjacent leave (sub-tree) diverge in their outputs upon receiving identical inputs. In this case, both the $\langle a \rangle$ and $\langle bbbbbb \rangle$ will be chosen to prompt LLMs. A more detailed process can be found in algorithm \ref{alg:ce_search}. This method ensures that selected queries not only exhibit similarity in terms of edit distance but also align in their properties, thereby teaching the LLM what is permissible and what is not in the language.

\begin{figure}[h]
  \centering
  \includegraphics[width=\linewidth]{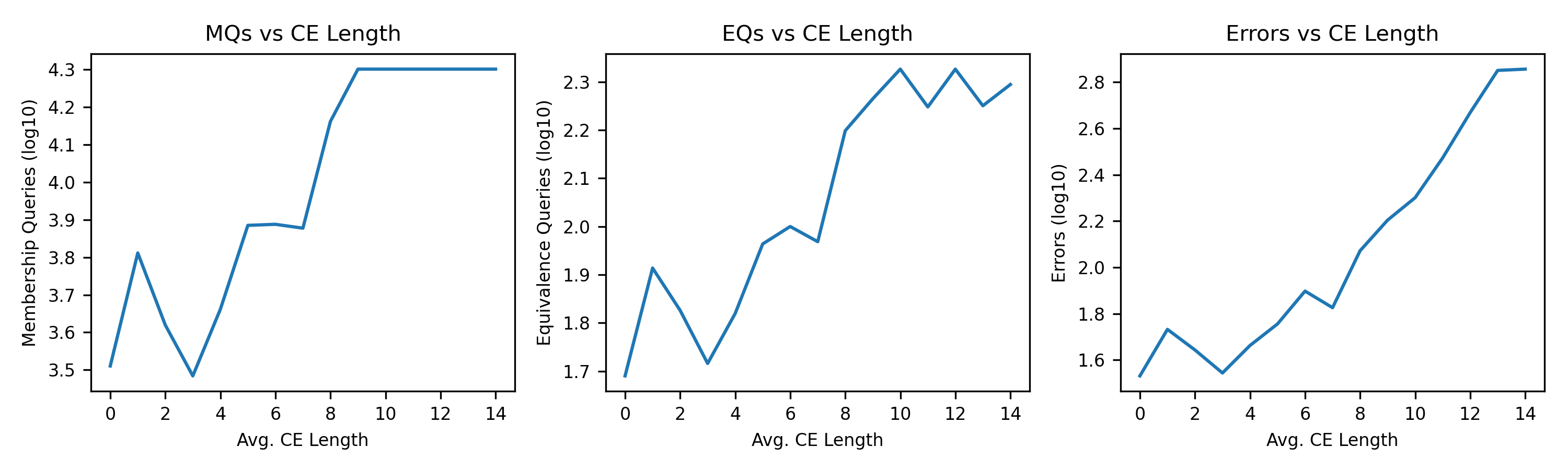}
  \caption{Relationship between errors and CE length. The average length of CEs refers to the length parameter used during generating CEs. They must be longer than this length parameter unless no longer counterexample exists. If the average CE length is set to 0, the oracle returns the shortest counterexample. This test is running on a simple automaton that recognizes strings with more than 3 'a's and more than 2 'b's, employing the \LEARNANYWAY{}  $\&$ TTT as the DFA learner.}
  \label{fig:errors_vs_ce_length}
\end{figure}

\subsection{\LEARNANYWAY{}  with Passive Refinement}
\citet{Angluin1997} investigate the issues of errors in membership queries and the polynomial-time learnability of DFAs using MQs and EQs). They demonstrate that DFAs can be learned in polynomial time despite malicious MQs. However, their method becomes impractical when the error rate in MQs grows exponentially with the average length of counterexamples.

Figure \ref{fig:errors_vs_ce_length} illustrates the relationship between MQs, EQs, and errors as the average length of counterexamples increases. Since active learning is driven by counterexamples, longer counterexamples extend the learning process. It is evident that the error count increases exponentially as the length of counter examples grows, even when the number of MQs and EQs plateaus.

The cost of asking tens of thousands of MQs to large language models (LLMs) for such a simple language is prohibitive. Moreover, when \LEARNANYWAY{}  cannot receive exactly same counterexamples to incorrect MQs, it fails to promptly correct persistent errors, causing error accumulation as new counterexamples are processed. Therefore, \LEARNANYWAY{}  struggles in environments where the shortest optimal examples are unavailable.

\begin{algorithm}[t]
\caption{\LEARNANYWAY{}  with Passive Refinement}\label{alg:LAPR}
\begin{algorithmic}
\State A dictionary that maps counter examples to MQs, $M$
\State Active Learner and passive learner, $L_a, L_p$
\State Expected epsilon, $\epsilon_e$
\Procedure{LAPR}{q}
\State Construct a hypothesis $\mathcal{H}$ from $L_a$
\While{True}
    \State $w_{ce}\leftarrow O_{EQ}(\mathcal{H})$
    \If{$w_{ce}=Null$}
        \State \textbf{return} $\mathcal{H}$
    \EndIf
    \State Add $w_{ce}$ to $C_{EQ}$ and update $C_{MQ}$ with its label
    \If{Hypothesis refinement $L_a(w_{ce})$ succeeds}
        \State Add all the membership queries during this step to $M$
    \Else
        \State Sample examples $Q^{+}, Q^{-}$ from $C_{EQ}$ and $C_{MQ}$
        \State Learn a hypothesis $\mathcal{H}_p$ from $Q^{+}, Q^{-}$
        \State Estimate $\epsilon$ using $\mathcal{H}_p$ and $C_{MQ}$
        \If{$\epsilon$ > $\epsilon_e$}
            \State Learn $\mathcal{H}_p$ by $L_a, O_{EQ}$ actively
        \EndIf
        \For{each $q_i$ in the $C_{MQ}$}
            \State $q_i.output\leftarrow\mathcal{H}_p(q_i)$
        \EndFor
        \State Resume $L_a$
    \EndIf
\EndWhile
\EndProcedure
\end{algorithmic}
\end{algorithm}

\paragraph{Algorithm.}
This algorithm, named \LEARNANYWAY{}  with Passive Refinement (LAPR), involves two distinct modes of learning: active learning and passive learning. It uses a pMAT, referred to as an oracle, which provides feedback through membership queries. We build our membership query process upon \LEARNANYWAY{} , it adds itself to the $C_{MQ}$ or replace the wrong query in $C_{MQ}$ with its label.
Initially, like other common active learning algorithms, the LAPR learner begins to hypothesize a model based on the input data it receives from the $O_{MQ}$. The active learner submits its hypothesis to the $O_{EQ}$, which either approves it if the hypothesis is equivalent to the target model or provides a counterexample. When a counterexample is received, it is recorded in a buffer that keeps track of MQs used during the analysis on equivalence queries. If the active learner successfully refines the hypothesis using the counterexample, the related membership queries during this refinement are stored in a dictionary for further use.

However, if the refinement based on the counterexample fails or the counterexamples accumulate beyond a practical limit, the algorithm shifts to a passive learning phase. In this phase, the algorithm samples data points from the previously collected MQs and EQs to form a new training set consisting of both positive and negative examples. The passive learner then constructs its own hypothesis $\mathcal{H}_p$ based on this new dataset, and use $\mathcal{H}_p$ to correct the related MQs. It also try to estimate the error rate ($\epsilon$), which measures the inconsistency between the outputs of the passive hypothesis and the outputs from the membership query cache. If this error rate exceeds a pre-determined threshold, indicating significant inaccuracies, the algorithm reverts to using the passive learning mode, but now guided more intensively by the oracle to refine the hypothesis.

Therefore, the algorithm acts more like classical active learning algoithms when $\epsilon$ is low. In contrast, the LAPR converge to passive learning algorithms when the $\epsilon$ is high. The overall goal of the LAPR algorithm is to continuously refine its understanding of the unknown language through a blend of active and passive learning techniques, making adjustments based on the feedback received from the oracle and the observed error rate, striving to develop an increasingly accurate representation of the target language.

\section{Empirical Results}

In this section, we report on a thorough analysis of the proposed prompts and algorithmic comparisons. we employed the implementations of active and passive learning algorithms available in LearnLib \cite{DBLP:conf/cav/IsbernerHS15}. This allowed for a standardized and fair comparison across different approaches. Our experiments were based on a dataset of DFAs derived from the 28 exercises provided in \cite{sipser1996introduction}, which offer a range of DFAs designed around introductory concepts in automata theory. These DFAs were chosen based on the assumption that if human learners can intuitively understand and construct DFAs, then LLMs should also possess the capability to effectively handle similar tasks.

The selected DFAs, while structurally simple, serve as an effective medium to assess the membership query answering capabilities of LLMs. 
This simplicity allows us to isolate and examine the effects of increasing the complexity of test scenarios by incrementally adjusting two main parameters in our experiments: the length of counterexamples and the error probability $\epsilon$ in the pMAT framework. 
Additionally, for each DFA, we provided one positive and one negative example, complete with detailed explanations as to why these examples are or are not part of the target language.

\subsection{Prompt Study}

\begin{table}[b]
\small
\centering
\caption{Performance Analysis of TTT \& LStar across Various Oracle Types}
\label{tab:prompt}
\begin{tabular}{lccccccc}
\toprule
\multirow{2}{*}{\textbf{Oracle Type}} & \multicolumn{2}{c}{\textbf{MQ}} & \multicolumn{2}{c}{\textbf{LLM Queries}} & \multicolumn{2}{c}{\textbf{Learning Acc.}} & \multirow{2}{*}{\textbf{Epsilon (samples)}}
\\ 
\cmidrule(lr){2-3} \cmidrule(lr){4-5} \cmidrule(lr){6-7}
~ & \textbf{TTT} & \textbf{LStar} & \textbf{TTT} & \textbf{LStar} & \textbf{TTT} & \textbf{LStar} & \\
\midrule
Target model        & 925 & 6393               & N/A                          & N/A                          & 1.0                  & 1.0                  & N/A                \\
\midrule
Baseline           & 225 & 266               & 174 & 204          & 0.107 & 0.071   & 0.156 (1390) \\
CoT                & 353 & 475               & 248 & 338          & 0.429 & 0.393   & 0.068 (1106) \\
Verification \& CoT (Ours)\textsuperscript{*}         & 451 & 522               & 309 & 413          & 0.464 & 0.429   & \textbf{0.044 (1040)} \\
Discriminator \& CoT (Ours)         & 492  & 714                & 357 & 561         & \textbf{0.571} & \textbf{0.536}  & 0.047 (1102)  \\
\bottomrule
\end{tabular}
\end{table}

The results depicted in Figure \ref{fig:errors_vs_ce_length} demonstrate that the Chain of Thought (CoT) methodology significantly enhances the accuracy of membership queries. When compared to a baseline prompt, which asks LLMs to answer yes or no, the introduction of CoT prompts reduces the error rate $\epsilon$ (calculated as the ratio of errors to total unique membership queries) to $0.068$. This improvement is attributed to CoT's capacity to generate more structured and contextually enriched responses from LLMs, thereby offering clearer guidance throughout the DFA learning process.

Further analysis reveals that the $\mathtt{Verification}$ prompt, which examines the alignment between the query inputs and the CoT responses and verify the valid usage of the target language definitions, considerably improves query accuracy. This prompt encourages LLMs to self-check and correct, which is crucial as it was observed that LLMs sometimes provided responses that were directly contradictory to their reasoning response. Incorporating $\mathtt{Verification}$ prompts reduced $\epsilon$ to $0.044$, marking the lowest error rate among the different prompting strategies tested.

Additionally, the $\mathtt{Discrimination}$ prompt, designed to utilize information from counterexamples to determine membership status, also significantly lowers $\epsilon$ to $0.047$. This prompt selects the most relevant counterexamples based on a discrimination tree and edit distance metrics. Notably, while the $\epsilon$ for the $\mathtt{Discrimination}$ prompt is not the lowest, it achieved the highest learning accuracy rates, achieving $0.571$ for TTT and $0.536$ for LStar, suggesting that this prompt effectively captures similarities between the inputs of membership queries and the counterexamples.

Additional finding is that the reduced number of membership queries correlates with decreased error rates in predictions made by the learned model, particularly in contexts where a probabilistic language model (LLM) serves as the oracle. The probabilistic nature of LLMs implies that fewer queries could reduce the cumulative likelihood of error propagation, thereby enhancing the overall quality of the learned automata. Our data suggest that TTT, by minimizing the number of queries, may thus be more effective at converging towards the correct automata structure.

\subsection{Algorithm Comparison}

We compared our LAPR algorithm with traditional DFA learning methods such as LStar and TTT, as well as the \LEARNANYWAY{}  algorithm, which handles incorrect membership queries through counterexamples. Our experimental setup involved restricting counterexample lengths to 5 and capping membership queries at 50 per DFA learning session, considering any excess as a timeout.

The comparative results, outlined in Table~\ref{tab:sample-table}, illustrate that while LStar and TTT falter with incorrect membership queries, \LEARNANYWAY{}  adjusts queries via counterexamples but struggles with complex DFAs due to the need for additional queries to analyze lengthy counterexamples. In contrast, LAPR consistently forms equivalent automata even without optimal counterexamples, employing a dual approach of using MQs and EQs to learn from and correct the persistent errors introduced by LLMs.

\begin{figure}[t]
  \centering
  \includegraphics[width=\linewidth]{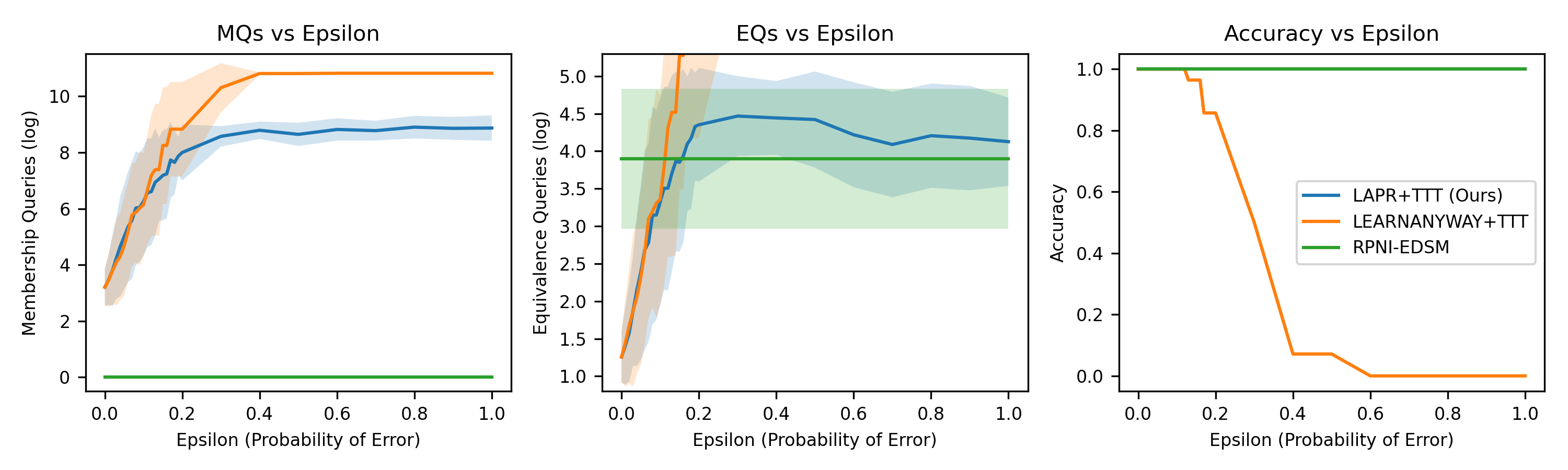}
  \caption{\LEARNANYWAY{} , RPNI-EDSM, LAPR Performance Comparison. Both RPNI-EDSM and LAPR can learn the correct DFAs within $5e4$ membership queries.}
  \label{fig:queries_vs_epsilon}
\end{figure}

To further validate LAPR's scalability and effectiveness, we introduced a probabilistic oracle scenario in our experiments, where the oracle might erroneously answer membership queries with a probability $\epsilon$, ranging from 0 to 1. Here, the counterexample length for equivalence queries was set to 10. As shown in Figure \ref{fig:queries_vs_epsilon}, LAPR required fewer MQs than \LEARNANYWAY{} , and unlike \LEARNANYWAY{} , which needed exponentially more counterexamples as $\epsilon$ increased, LAPR maintained a stable number of required counterexamples, comparable to those needed by RPNI-EDSM.

An interesting observation was that with $\epsilon$ greater than 0.18, passive algorithms like RPNI-EDSM proved more efficient than active learning methods due to their reliance solely on counterexamples. However, as shown in Table~\ref{tab:prompt}, proper prompting can reduce $\epsilon$ to below 0.1 when using LLMs as oracles, making our LAPR method more effective than both \LEARNANYWAY{}  and RPNI-EDSM in such settings.

\begin{table}[ht]
  \centering
  \caption{Performance analysis of various algorithms across different oracle types}
  \label{tab:sample-table}
  \begin{tabular}{lcccc}
    \toprule
    \multirow{2}{*}{Algorithm} & \multicolumn{4}{c}{Oracle} \\
    \cmidrule(lr){2-5}
    & CoT & CoT \& Verification & CoT \& Discrimination & Best \\
    \midrule
    LStar & 0.286 & 0.333 & 0.667 & 0.667 \\
    TTT & 0.333 & 0.393 & 0.571 & 0.571 \\
    \LEARNANYWAY{}  \& TTT & 0.833 & 0.833 & 0.833 & 0.833 \\
    LAPR \& TTT & \textbf{1.0} & \textbf{1.0} & \textbf{1.0} & \textbf{1.0} \\
    \bottomrule
  \end{tabular}
\end{table}

\subsection{Discussion, Limitations, and Challenges}

Despite the advances made with our LEARNANYWAY with Passive Refinement (LAPR) algorithm, there remains scope for enhancement, particularly in terms of the algorithm's efficiency in halting execution when faced with low-quality queries. Critical to this improvement is the ability to more precisely estimate $\epsilon$, the error rate, and to detect conflicts with greater accuracy. These enhancements necessitate an improved design of the passive learner component.

A potential challenge arises from the trade-off between building a proper dataset, including all queries from $C_{MQ}$ and $C_{EQ}$, and another approach that focuses solely on counterexamples. Utilizing a complete dataset tends to introduce noise due to the erroneous data stemming from the LLMs' responses. This situation can lead to a critic that does not accurately reflect the target automaton's structure due to the high incidence of errors. On the other hand, constraining the training data to only include counterexamples ensures the correctness of the training inputs but results in a significantly smaller dataset. This limited dataset can cause the passive learner to overfit to these examples, making it less generalizable and poorly equipped to handle new, unseen queries effectively. This situation underscores the need for a balanced approach in training the passive critic, where both the integrity of training data and the volume of data are optimized to enhance the overall performance of LAPR.

\section{Conclusion}

This paper explored the integration of Large Language Models (LLMs) into the domain of automata learning, culminating in the development of the probabilistic Minimally Adequate Teacher (pMAT) framework. This approach capitalizes on the probabilistic and sometimes erroneous outputs of LLMs to answer membership queries, supplemented by accurate counterexamples for hypothesis refinement.

Our key contributions include the introduction of innovative prompting strategies—the $\mathtt{Discrimination}$ and $\mathtt{Verification}$ prompts, which significantly enhance the reasoning accuracy of LLMs when responding to membership queries. Furthermore, we implemented and evaluated the \LEARNANYWAY{}  with Passive Refinement (LAPR) algorithm, which adeptly combines active and passive learning techniques. This hybrid approach allows for effective management of the persistent errors typical in LLM outputs, thereby facilitating the construction of accurate automata even when faced with high error rates in membership queries. The empirical results underscore the robustness and efficiency of LAPR, particularly in comparison to traditional DFA learning algorithms like LStar and TTT, and other methods such as \LEARNANYWAY{}  that rely solely on active learners.

Through rigorous empirical evaluation, we demonstrated that our \LEARNANYWAY{}  with Passive Refinement (LAPR) algorithm outperforms traditional DFA learning algorithms such as LStar and TTT, particularly in environments where counterexamples may be lengthy or suboptimal. LAPR's ability to dynamically switch between active and passive learning phases allows for efficient error correction and hypothesis refinement, effectively handling the exponential increase in persistent errors associated with higher counterexample lengths. This approach mitigates the inherent limitations of using LLMs as oracles and enhances the potential for developing accurate and reliable automata for a wide range of applications in computational linguistics, software engineering, and beyond.


\bibliographystyle{plainnat}
\bibliography{neurips_2024}

\newpage

\appendix

\section{Appendix / supplemental material}

\subsection{Verification Prompt Example}

\begin{figure}[h]
  \centering
  \includegraphics[width=\linewidth]{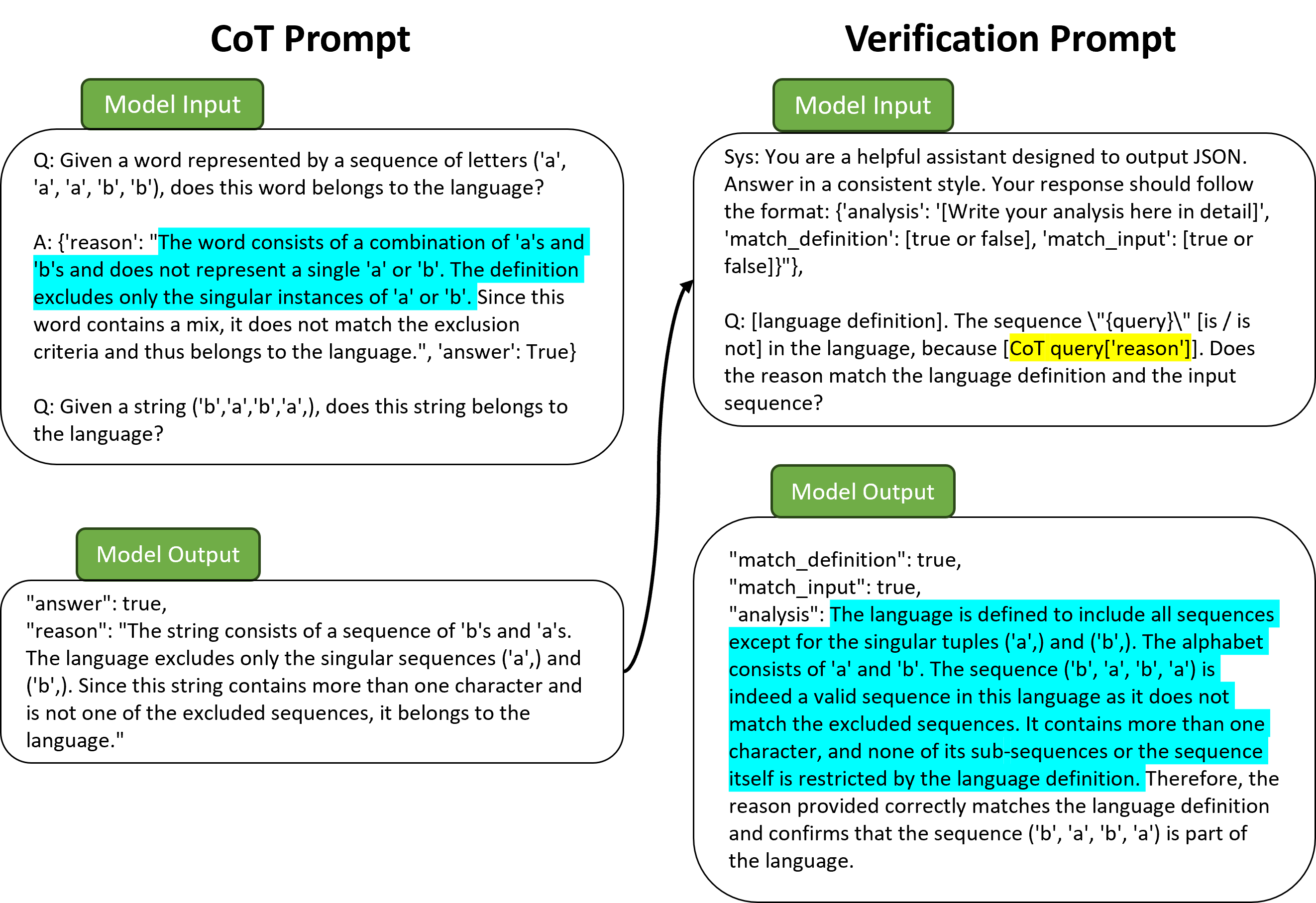}
  \caption{Verfication prompt}
  \label{fig:veri_prompt}
\end{figure}

\newpage

\subsection{Discrimination Prompt Words Selecting Algorithm}

\begin{algorithm}
\caption{Find Similar Counterexamples by Discrimination Tree}
\label{alg:ce_search}
\begin{algorithmic}
\Procedure{HistoryOracle}{C,q}
    \If{$q\in C_{EQ}$}
        \State \textbf{return} $C_{EQ}[q]$
    \Else
        \State \textbf{return} False
    \EndIf
\EndProcedure
\State Represent \textit{HistoryOracle} by $O_{\mathcal{H}}$
\Procedure{Discrimination Based Word Search}{q}
\State Construct the hypothesis $\mathcal{H}$ via any active learner by $O_{\mathcal{H}}$
\State Build the discrimination tree $DT$ based on $\mathcal{H}$
\State Find the lowest common ancestor $d$ of $\mathcal{H}[q]$ and another child $s$ of $d$.
\State Use \textit{Levenshtein Distance} $L$ to estimate the similarity between two words.
\State Initialize $ l_q\leftarrow \inf, l_s\leftarrow \inf$.
\State Let $L$ represent Levenshtein Distance.
\For{each query $q_i$ in the $C$}
    \If{$\mathcal{H}[q_i]=\mathcal{H}[q]$ and $L(q_i, q) < l_q$}
        \State $l_q\leftarrow L(q_i, q)$
        \State $w_q\leftarrow q_i$
    \ElsIf{$\mathcal{H}[q_i]=\mathcal{H}[s]$ and $L(q_i, q) < l_s$}
        \State $l_s\leftarrow L(q_i, q)$
        \State $w_s\leftarrow q_i$
    \EndIf
\EndFor 
\State \textbf{return} $w_q, w_s$
\EndProcedure
\end{algorithmic}
\end{algorithm}

\end{document}